# A Novel Adaptive Routing through Fitness Function Estimation Technique with Multiple QoS Parameters Compliance

Dr. T. R. Gopalakrishnan Nair *, Kavitha Sooda**

- Director, Research Industry and Incubation Centre, Dayananda Sagar Institutions, Bangalore- 560 078, India, Member, IEEE
trgnair@yahoo.com

**Asst. Prof, Department of CSE, NMIT, Bangalore- 560 064, India
kavithasooda@gmail.com

**Abstract :** The paper presents a method which shows a significant improvement in discovering the path over the distance vector protocol. The proposed method is a multi-parameter QoS along with the fitness function which shows that it overcomes the limitation of DV like routing loops by spanning tree approach, count-to-infinity problem by decision attribute. The input considered is a topology satisfying the QoS parameters of size 1 to 64 nodes and it was shown that an optimal path selection was obtained efficiently over the classical distance vector algorithm.

**Keywords :** Distance Vector algorithm, fitness function estimation method, QoS, optimal path, routing.

## 1 INTRODUCTION

Technology for routing is the most challenging aspects for achieving efficient networks. Many ad hoc routing technique review strategies and heuristic approaches have been applied to overcome some of the issues related to routing, which has shown comparatively optimal solution to achieve the desired results [1],[2]. Adaptive routing approaches were discussed in [3] and till date lot of proposals have been put forth for the enhancement of dynamic routing.

There is a dire need of intelligent approaches for dealing with routing with considerable enumeration of parameters for optimal path selection.

This paper deals with bandwidth, delay loss and jitter together with intelligence.

In this short paper we have four sections. Section 2 gives the simulation results and discussion on the obtained results and finally conclusion is given in section 3.

## 2 SIMULATION RESULTS

Table 1, shows one such result of the dynamic creation of the topology. The result obtained after applying distance vector algorithm, on the topology created at random, with minimum number of hops.

Table. 1. Path taken by DV approach

| Distance vector Algorithm | Source | Destination | Hop count | Path |
|---|---|---|---|---|
| | 1 | 0 | 1 | 1->0 |
| | 0 | 3 | 1 | 0->3 |
| | 4 | 1 | 5 | 4→7->3->0->2->1 |
| | 5 | 10 | 3 | 5->7->11->10 |
| | 9 | 19 | 3 | 9->12->20->19 |
| | 2 | 55 | 11 | 2->10->12->20->28->25->33->37->45->53->48->55 |

Similarly, Table 2, shows the result obtained after applying fitness function estimation algorithm, on the topology created at random, with minimum number of hops. We see that the fitness function estimation results show more powerful results, as QoS parameters are taken into consideration. It was observed that if sufficient bandwidth was not available then the route could not be estimated in our approach, but the classical distance vector approach shows some path through which data cannot flow.

Table. 2. Path taken by fitness function estimation

| Fitness function estimation result | Source | Destination | Hop count | Path |
|---|---|---|---|---|
| | 1 | 0 | 1 | 1->0 |
| | 0 | 3 | - | No sufficient bandwidth available |
| | 4 | 1 | 5 | 4→7->3->0->2->1 |
| | 5 | 10 | 1 | 5->10 |
| | 9 | 19 | 3 | 9->12->20->19 |
| | 2 | 55 | 10 | 2->0->1->9->17->25->33->37->45->53->55 |

The strength of the method lies in the bandwidth availability. Otherwise the method equally performs well to DV and gives the hop count lesser than or equal to the DV hop count. The following graph shows the graph for the same.

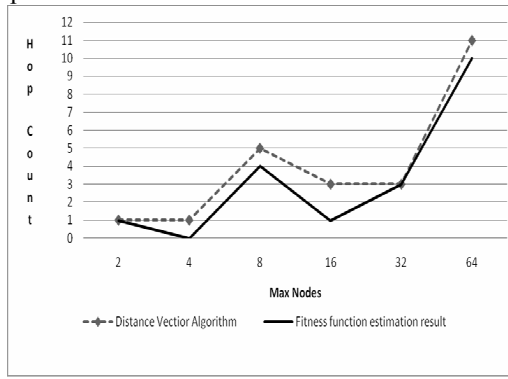

Fig. 1. Comparison of the path taken by the DV and Fitness Function Estimation approach.

## 3 CONCLUSION

This short paper presented a multi-parameter selection method based on forward channel scheme. This algorithm shows 100% assurance on bandwidth. However, overall optimization of QoS including bandwidth shows a promising approach over the distance vector algorithm. The looping problem was successfully overcome by the spanning tree approach and the count-to-infinity problem by the decision attribute of the fitness function which was simulated on a total of 64 nodes.

The approach can be extended with application specific and priority based, so as to make the routing approach faster.